\documentclass[10pt,a4paper]{article}
\usepackage{geometry}

\usepackage[utf8]{inputenc}

\usepackage{nameref,hyperref}

\usepackage{changepage}

\usepackage[aboveskip=1pt,labelfont=bf,labelsep=period,singlelinecheck=off]{caption}

\makeatletter
\renewcommand{\@biblabel}[1]{\quad#1.}
\makeatother

\usepackage{lastpage,fancyhdr,graphicx}
\usepackage{epstopdf}
\fancyheadoffset[L]{2.25in}
\fancyfootoffset[L]{2.25in}

\usepackage{color}

\usepackage{graphicx}

\newcommand{\vect}[1]{\mathbf{#1}}

\begin{document}
\vspace*{0.35in}
\begin{flushleft}
{\Large
\textbf\newline{Electric-field control of spin-orbit torques in perpendicularly magnetized W/CoFeB/MgO films}
}
\newline
\\
Mariia\,Filianina\textsuperscript{1,2},
Jan-Philipp\,Hanke\textsuperscript{1,3},
Kyujoon\,Lee\textsuperscript{1},
Dong-Soo\,Han\textsuperscript{1,4},
Samridh\,Jaiswal\textsuperscript{1,5},
Gerhard\,Jakob\textsuperscript{1},
Yuriy\,Mokrousov\textsuperscript{1,3}
Mathias\,Kl\"aui\textsuperscript{1,2*}
\\
\bigskip
\bf{1} Institute of Physics, Johannes Gutenberg University, Mainz, Germany
\\
\bf{2} Graduate School of Excellence Material Science in Mainz, Germany
\\
\bf{3} Peter Gr{\"u}nberg Institut and Institute for Advanced Simulation, Forschungszentrum J\"ulich and JARA, J{\"u}lich, Germany
\\
\bf{4} Center for Spintronics, Korea Institute for Science and Technology, Seoul, Republic of Korea
\\
\bf{5} Singulus Technology AG, Kahl am Main, Germany
\bigskip
\\
* klaeui@uni-mainz.de

\end{flushleft}

\date{\today}

\section*{Abstract}
Controlling magnetism by electric fields offers a highly attractive perspective for designing future generations of energy-efficient information technologies. Here, we demonstrate that the magnitude of current-induced spin-orbit torques in thin perpendicularly magnetized CoFeB films can be tuned and even increased by electric field generated piezoelectric strain. Using theoretical calculations, we uncover that the subtle interplay of spin-orbit coupling, crystal symmetry, and orbital polarization is at the core of the observed strain dependence of spin-orbit torques. Our results open a path to integrating two energy efficient spin manipulation approaches, the electric field-induced strain and the current-induced magnetization switching, thereby enabling novel device concepts.
\bigskip


Controlling efficiently the magnetization of nanoscale devices is essential for many applications in spintronics, and is, thus,  attracting significant attention in basic and applied science.
In recent years, current-induced switching via 
spin-orbit torques (SOTs)~\cite{Manchon_review} has emerged as one of the most promising approaches to realize scalable magnetoresistive random-access memories (MRAM).
 The SOT-induced switching is realized in a ferromagnet/heavy metal (FM/HM) bilayers, where the existence of sizable damping-like $ \vect{T}^{||} 
\propto \vect{m} \times (\vect{y} \times \vect{m})$ and field-like $\vect{T}^{\bot} \propto 
\vect{m} \times \vect{y}$ components of the SOT due to the flow of an electrical current along the $x$-direction was theoretically and experimentally studied.\cite{Manchon_prb_2008,Miron_nmat_2010,Kim_nmat_2012, Garello_nnano_2013,Liu_science,Schulz_PRB_SOT,Schulz_apl_sot, LoConte_APL} These torques originate from the spin Hall effect in the bulk of the HM material~\cite{Sinova_RMP_2015} and the inverse spin galvanic effect at the FM/HM interface.~\cite{Belkov_Ganichev}  

It was shown that the damping-like torque term can be large enough to switch the magnetization direction at low current densities down to $10^7-10^8$ A cm$^{-2}$,~\cite{Miron_nature_2011, Liu_science} which makes them particularly attractive for device applications.~\cite{Prenat_2016} 

While sample parameters such as composition and layer thickness of FM/HM heterostructures can be adjusted to design the magnitude and the sign of SOTs, their ``dynamical'' control in a given system on-demand by external means is of great fundamental and technological interest. One of the energy-efficient tools for that is offered by the use of electric field-induced mechanical strain.\cite{Wang_strain_review} 
Avoiding the need for electrical currents and, thus, eliminating the associated losses, strain is known to effectively tune magnetic properties such as magnetic anisotropy and, consequently, the magnetic domain structure and dynamics of in-plane thin films.~\cite{Sohn_ACS_nano_2015,Finizio_PRA,Filianina_APL, Foerster_2017} Moreover, as strain can be applied locally, it provides a playground to develop and realize complex switching concepts in simplified device architectures.

While attempts were made to investigate the effect of strain on switching by spin torques,~\cite{Wang_PRA_2018, Huang_ACMP_2016,Nan_AFM_2019} primarily the effect of strain on the anisotropy and the resulting impact on the switching was studied. Furthermore, these previous studies focused exclusively on systems with in-plane magnetic easy axis and experimental studies in perpendicularly magnetized multilayers are still elusive.
However, in the light of the potential for technological applications, it is most desirable to optimize all magnetic parameters including the SOTs in ferromagnetic elements. In particular using systems with perpendicular magnetic anisotropy (PMA) is attractive
as increased thermal stability, higher packing densities and improved scaling behavior are intrinsic to PMA materials as compared to their in-plane magnetized counterparts.~\cite{Nishimura_JAP, Ikeda_nmat_2010}

In this work, we demonstrate electrically induced strain control of SOTs in perpendicularly magnetized W/CoFeB/MgO multilayers grown on a piezoelectric substrate. The SOTs are evaluated by magnetotransport and second-harmonic methods under in-plane strain of different character and magnitudes. We find that the strain, as modulated by the electric field applied across the piezoelectric substrate, leads to distinct responses of field-like and damping-like torques, with a large change of the latter by a factor of two. Based on the electronic structure of realistic heterostructures, we explain our experimental findings by theoretical \textit{ab initio} calculations and reveal the microscopic origin of the observed strain effects on the magneto-electric coupling and the spin-orbit torques.

\begin{figure}
\centering
\includegraphics[width=0.6\columnwidth]{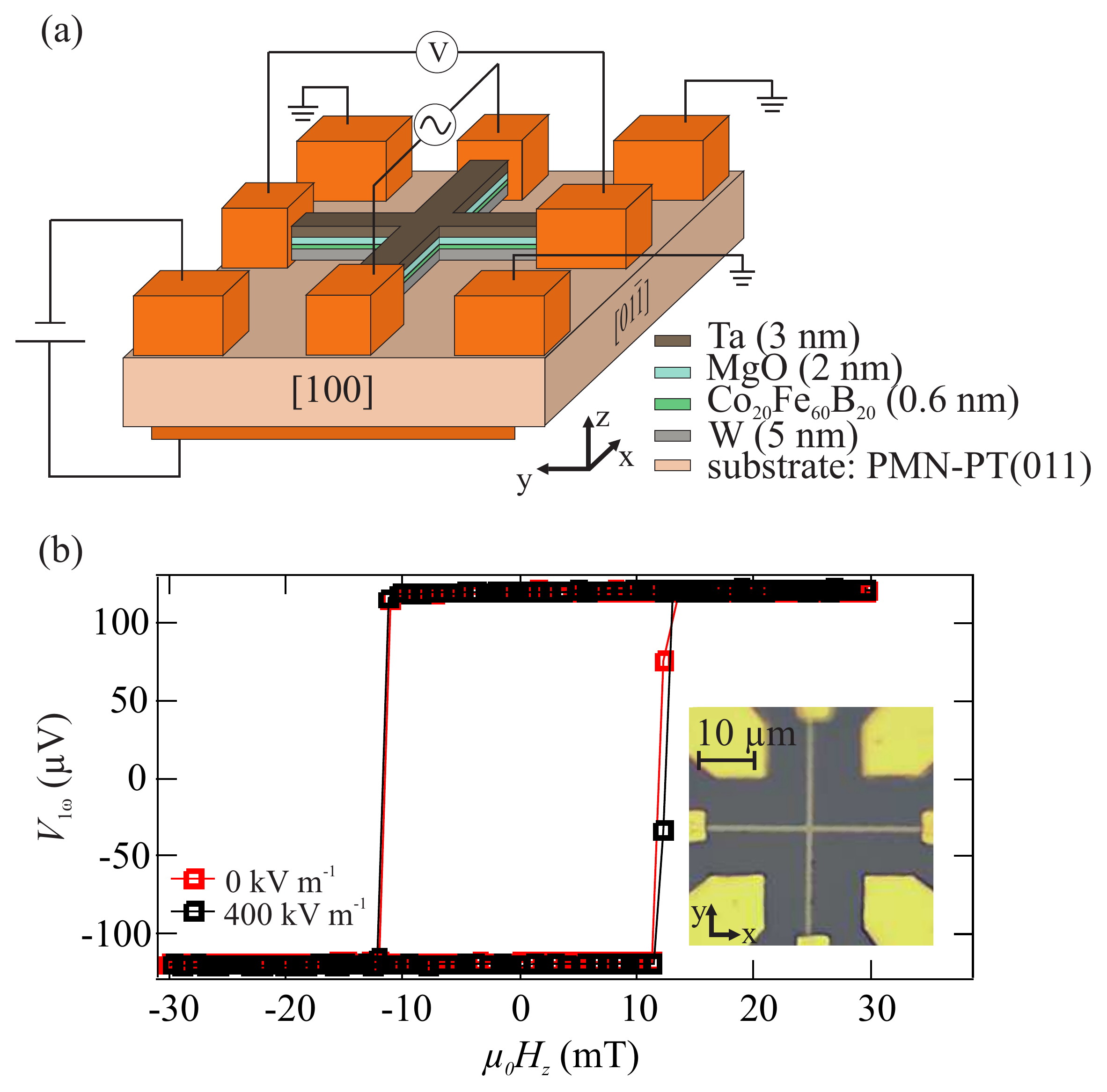}
\caption{(a) Schematic of the Hall-cross device fabricated on top of the PMN-PT$(011)$ substrate and the electrical contacts used for the application of the OOP electric field to generate strain as well as the electrical contacts to the Hall bar. In this configuration the current flow ($x$-axis) is along the $[01\bar{1}]$ direction of the PMN-PT substrate, thus, in the text it is referred to as tensile strain configuration. For compressive strain, the the current flow ($x$-axis) is along the $[100]$ direction. (b)~$1\omega$ Hall voltage hysteresis loop measured in the OOP direction at $0$ kV m$^{-1}$ (red), and $400$ kV m$^{-1}$ (black) applied to the PMN-PT substrate. The inset shows the optical microscope image of the Hall-cross structure used for the spin torque measurements.   
}
 \label{sample_environment}
\end{figure}

Figure~\ref{sample_environment} (a) shows the schematic of the Hall-cross device employed for the measurements of the damping-like (DL) and the field-like (FL) effective SOT fields in W($5$ nm)/CoFeB($0.6$ nm)/MgO($2$ nm)/Ta($3$ nm) multilayer fabricated on a [Pb(Mg$_{0.33}$Nb$_{0.66}$O$_3$)]$_{0.68}$-[PbTiO$_3$]$_{0.32}(011)$ (PMN-PT) substrate, employed to electrically generate mechanical strain. An optical microscope image of the Hall-cross device used in the experiment is presented in the inset in Fig.~\ref{sample_environment} (b) and more details are provided in the Methods section.



Uniaxial in-plane strain was generated by applying an out-of-plane (OOP) DC electric field across the piezoelectric PMN-PT$(011)$ substrate. The piezoelectric strain response to the applied electric field exhibits a hysteretic behavior.~\cite{Wu_JAP_2011} However, electric fields that exceed the material-specific coercive field pole the substrate and promote a regime where the generated strain is characterized by a linear response. The linear regime is maintained until the substrate is poled in the other direction by application of the electric fields larger than the opposite coercive field.~\cite{Wu_JAP_2011} 
Therefore, before the first measurements, but after the structuring process, we poled the PMN-PT substrate by applying an electric field of $+400$ kV m$^{-1}$. In the following, we used the DC electric fields that allowed us to vary the strain within the linear response regime,~\cite{Wu_JAP_2011} as this provides reliable electrical control over the induced strain.

We also note, that the Hall cross in Fig.~\ref{sample_environment} (b) was fabricated such that the arms were oriented along the $[01\bar1]$ and $[100]$ directions of the PMN-PT$(011)$ substrate, which correspond to the directions of tensile and compressive strain, respectively, as set by the crystallographic structure of the substrate.~\cite{Wu_JAP_2011}
The experimental results of the SOTs obtained in the configuration with the current ($x$-axis) flowing along the $[01\bar1]$ and $[100]$ directions will be referred to as modified by tensile and compressive strain, respectively [Fig.~\ref{sample_environment} (a)].

First, we characterize the magnetic hysteresis of the system at zero DC electric field. Figure~\ref{sample_environment} (b) shows the anomalous Hall voltage sweep with the OOP magnetic field ($\mu_0H_z$) measured for W/CoFeB/MgO/Ta at $0$ kV m$^{-1}$ (red line), demonstrating the easy-axis switching typical for W-based thin CoFeB stacks.~\cite{Jaiswal_APL, Takeuchi_torques} The OOP magnetization loop, measured at $400$ kV m$^{-1}$ (black line), is overlaid on top of it and shows no sizeable change due to the generated strain indicating that the system has always a dominating PMA.

The current-induced effective SOT fields were measured using 2$\omega$ Hall measurements~\cite{Hayashi_PRB, Pi_SOT} (see Methods section for more details).

\begin{figure}
\centering
\includegraphics[width=0.7\columnwidth]{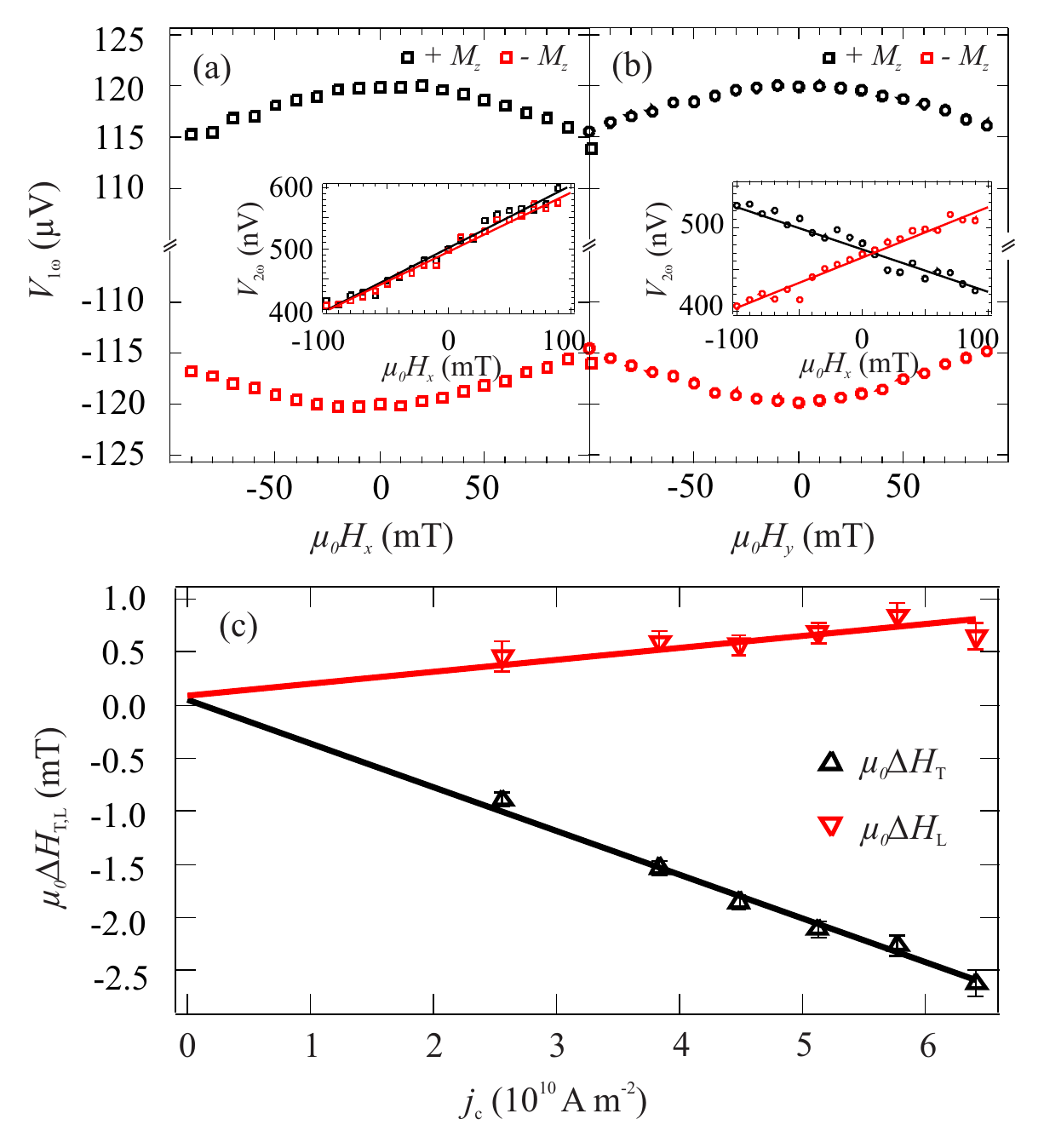}
\caption{(a)~$V_{1\omega}$ and $V_{2\omega}$ (inset) signals as a function of the in-plane field directed along the current flow. (b)~$V_{1\omega}$ and $V_{2\omega}$ (inset) signals as a function of the in-plane field directed transverse to the current flow. The data were measured at the current density of $3.8 \times 10^{10}$ A m$^{-2}$. Black and red symbols represent signals for the magnetization pointing along $+z$ and $-z$, respectively. (c)~The longitudinal ($\mu_0\Delta H_\mathrm{L}$) and the transverse ($\mu_0\Delta H_\mathrm{T}$) components of the SOT effective field plotted as a function of current density $j_\mathrm{c}$. At each value of current density, the averaged values of the SOT effective field for $+M_z$ and $-M_z$ are shown. 
}
 \label{SOT_data}
\end{figure}

Fig.~\ref{SOT_data} shows the representative in-plane field dependencies of the first ($V_{1\omega}$) and the second ($V_{2\omega}$) harmonics of the Hall voltage when an AC current with the current density of $j_\mathrm{c}=3.8 \times 10^{10}$~A m$^{-2}$ was applied to the current line. The DC poling voltage was set to zero, thus, no strain was imposed on the Hall cross. The longitudinal [Fig.~\ref{SOT_data} (a)] and the transverse [Fig.~\ref{SOT_data} (b)] field sweeps exhibit the expected symmetries: for the longitudinal field, the slopes of $V_{2\omega}$ versus the field are the same for both magnetization directions along $+z$ ($+M_z$) or $-z$ ($-M_z$), whereas their sign reverses for the transverse field sweep.

Using the procedure described in Methods section we analyze the transverse ($\mu_0\Delta H_\mathrm{T}$) and the longitudinal ($\mu_0\Delta H_\mathrm{L}$) components of the SOT effective field for both magnetization directions $\pm M_z$ and plot the average of these field components as a function of the applied current density $j_\mathrm{c}$ in Fig.~\ref{SOT_data} (c). The resulting linear dependencies are fitted such that the slopes $ \mu_0\Delta H_\mathrm{T}/j_\mathrm{c}$ and $ \mu_0\Delta H_\mathrm{L}/j_\mathrm{c}$ determine the FL, $\mu_0H^\mathrm{eff}_\mathrm{{FL}}$, and the DL, $\mu_0H^\mathrm{eff}_\mathrm{{DL}}$, SOT effective fields, respectively. Similarly, the effective field were extracted for different DC electric fields applied to the PMN-PT substrate to vary the magnitude of the generated strain. 

\begin{figure}
\centering
\includegraphics[width=0.6\columnwidth]{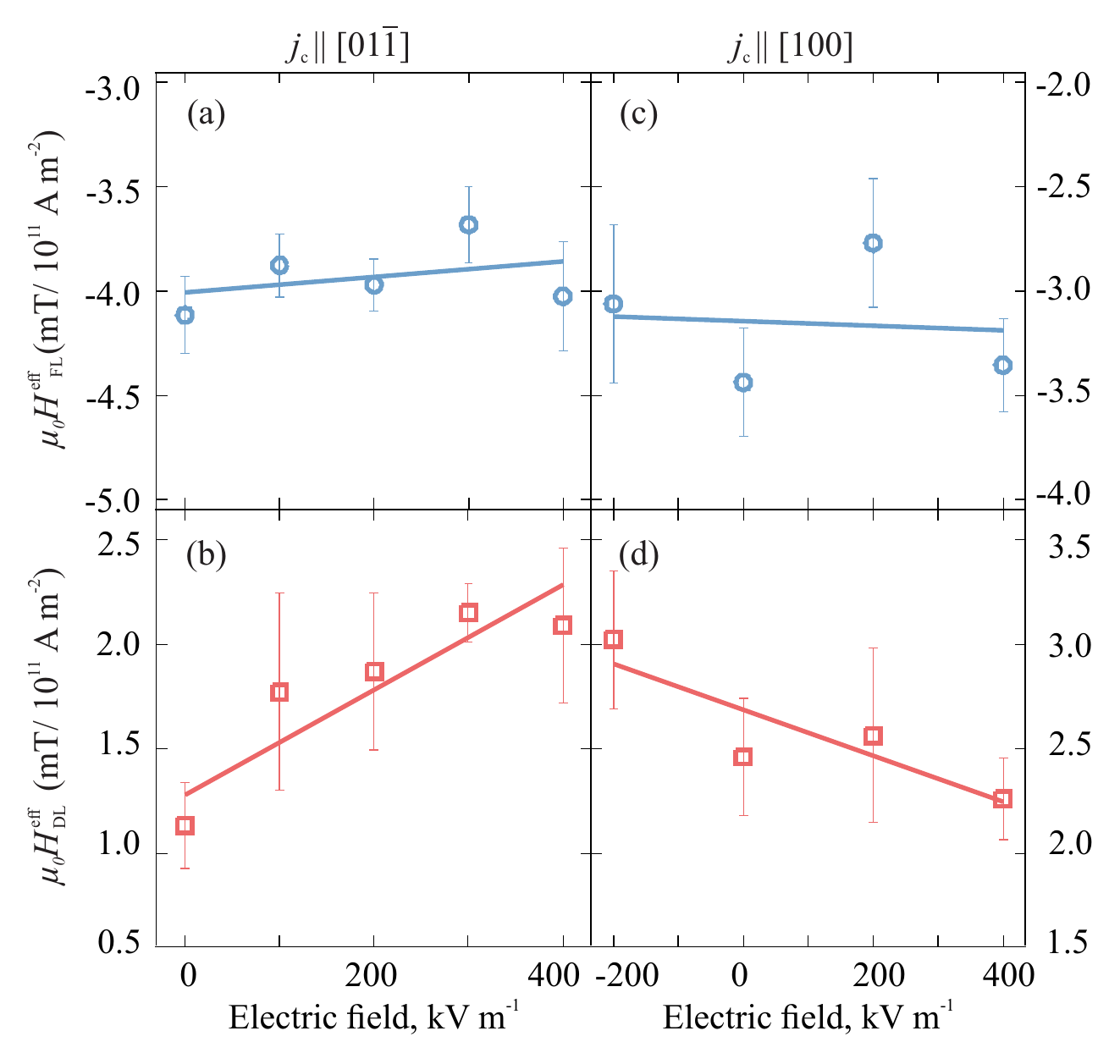}
\caption{ (a)~FL and (b)~DL SOT effective fields as a function of the electric field applied across the PMN-PT$(011)$ for the current flowing along the tensile ($[01\bar{1}]$) strain direction. (c)~FL and (d)~DL SOT effective fields as a function of the electric field applied for the current flowing along the compressive ($[100]$) strain direction. The solid lines represent the linear fit of the data to guide the eye.   
}
 \label{SOT_strain}
\end{figure}

The electric field dependent results are summarized in Fig.~\ref{SOT_strain}. We obtained that the FL torque does not change significantly for both tensile and compressive strains as shown in Figs. \ref{SOT_strain} (a) and (c). On the contrary, Fig.~\ref{SOT_strain} (b) demonstrates that the tensile strain increases the DL torque up to two times when $400$~kV m$^{-1}$ is applied, which corresponds to ca. $0.07 \%$ strain.\cite{Wu_JAP_2011} 
On the other hand, when the current is flowing along the compressive strain direction, the magnitude of the DL torque decreases with increasing strain. Thus, we find experimentally that the magnitude of the DL torque increases (decreases) upon the application of electrically induced tensile (compressive) strain. 

\begin{figure}
\centering
\includegraphics[width=\columnwidth]{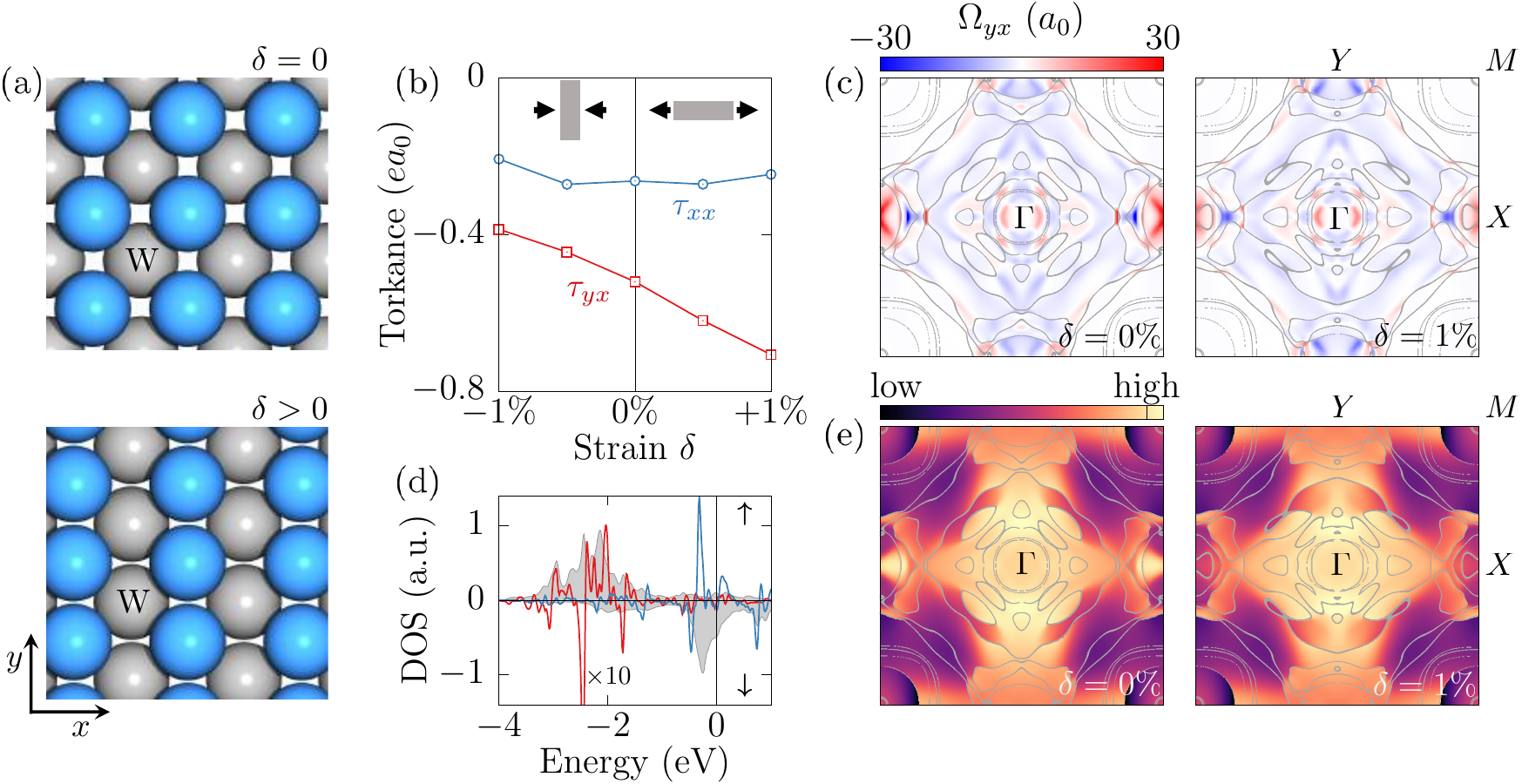}
\caption{ (a)~The uniaxial strain $\delta$ modifies the equilibrium crystal structure of the Fe$_{0.7}$Co$_{0.3}$/W(001) film, and reduces the symmetry from $C_{4v}$ to $C_{2v}$. (b)~Dependence of FL (blue) and DL (red) torkances on strain along the direction of the electric field, where a constant broadening of the energy bands by $25\,$meV is used. (c)~Microscopic contribution of all occupied bands to the DL SOT in relaxed and strained crystal structure. Gray lines indicate the Fermi surface. (d)~As compared to the behavior without strain (gray), the density of $d_{yz}$-states in the magnetic layer changes for majority (red) and minority (blue) spin channels owing to tensile strain of $\delta=1\%$. The red and blue curves, showing the difference with respect to the unstrained case, are scaled by a factor of $10$. (e)~Momentum-space distribution of the $d_{yz}$-polarization of all occupied majority states in the magnetic layer of the relaxed and strained system.}
\label{fig:1}
\end{figure}

 In order to understand the microscopic origin of the experimentally observed strain dependence of FL and DL SOTs, we perform density functional theory calculations of the electronic structure of Fe$_{1-x}$Co$_x$/W$(001)$, which consists of a perpendicularly magnetized monolayer and non-magnetic underlayers (see Methods section).
As illustrated in Figure~\ref{fig:1} (a), we expand or contract the crystal structure while keeping the in-plane area of the unit cell constant to account for the effect of uniaxial strain. This strain can be quantified by the ratio $\delta=(a_j'-a_j)/a_j$, where $a_j$ and $a'_j$ denote the lattice constant along the $j$th in-plane direction in the relaxed and distorted case, respectively. As a consequence, any finite strain reduces the original $C_{4v}$ crystal symmetry to $C_{2v}$, see Fig.~\ref{fig:1} (a). We employ a Kubo formalism~\cite{Freimuth2014a} to represent the SOT $T_i = \tau_{ij} E_j$ acting on the magnetization as the linear response to the applied electric field $E_j$, mediated by the torkance $\tau_{ij}$.  Owing to the mirror symmetries of the strained films with OOP magnetization, the torkances $\tau_{xx}$ and $\tau_{yy}$ characterize FL SOTs rooted in the electronic structure at the Fermi surface, whereas $\tau_{xy}$ and $\tau_{yx}$ correspond to DL torques, to which also electrons of the Fermi sea contribute. In order to model additionally disorder and temperature effects, we evaluate these response coefficients using a constant broadening $\Gamma=25\,$meV of the first-principles energy bands.~\cite{Freimuth2014a} In the following, $\delta$ refers to the strain along the orientation of the applied electric field, which points into $x$-direction.

Based on our electronic-structure calculations, we obtain the $\delta$-dependence of the SOTs shown in Fig.~\ref{fig:1} (b), which reveals similar qualitative trends as found in the experiment. Since FL and DL SOTs originate from different electronic states, they generally follow distinct dependencies on structural details. Specifically, while the FL term $\tau_{xx}$ is hardly affected if $\delta$ is varied, we predict that the magnitude of the DL torque $\tau_{xy}$ increases (decreases) linearly with respect to tensile (compressive) strain. For instance, expanding the lattice by $1\%$ along the electric-field direction drastically enhances the DL torkance by about $35\%$. To elucidate this remarkable behavior, we compare in Fig.~\ref{fig:1} (c) the momentum-space distribution of the microscopic contributions to the DL SOT for relaxed and strained films. In contrast to the occupied states around the $M$-point that are barely important, electronic states near the high-symmetry points $\Gamma$, $X$, and $Y$ constitute the major source of the DL torkance. In particular, tensile strain promotes strong negative contributions around $X$ and $Y$ [see Fig.~\ref{fig:1} (c)], leading to an overall increase in the magnitude of $\tau_{yx}$ as depicted in Fig.~\ref{fig:1} (b).

To further associate our findings with the underlying electronic structure, we turn to the orbital polarization of the states in the magnetic layer, the physics of which is dominated by $d$-electrons. Whereas the behavior of $d_{xy}$, $d_{x^2-y^2}$, and $d_{z^2}$ is independent of the sign of the applied strain $\delta$, the states of $d_{yz}$- and $d_{zx}$-character transform manifestly differently with respect to tensile or compressive strain. Remarkably, the latter orbitals also mediate the hybridization with the heavy-metal substrate, which implies that their dependence on structural details offers additional microscopic insights into the SOTs in the studied thin films. As an example, we consider in Fig.~\ref{fig:1} (d) the strain-induced change of the density of $d_{yz}$-states in the magnetic layer as compared to the case with four-fold rotational symmetry. While the density of minority-spin states at the Fermi level is hardly affected by tensile strain, the majority-spin states are redistributed rather strongly. As revealed by the momentum-resolved orbital polarization in Fig.~\ref{fig:1} (e), microscopically, this effect stems from pronounced $\delta$-driven variations of the $d_{yz}$-polarization around the $X$-point, which correlates with the presented changes of the DL torkance, Fig.~\ref{fig:1} (c).

As the system considered in this work has a relatively strong PMA, the static magnetic properties of the CoFeB film as visible from the hysteresis loop (Fig.~\ref{sample_environment}b) did not show any significant change with the applied strain. 
Prior work has focused on systems where the dominating effect of the strain was a change of the anisotropy,\cite{Wang_PRA_2018, Huang_ACMP_2016,Nan_AFM_2019} but here we have strong PMA and probe the change of the SOTs as the main factor. The sizable change in the torques found can be explained by our theoretical calculations.


Using our microscopic insights obtained from the electronic structure calculations, we uncovered that the distinct nature of the experimentally observed trends for FL and DL torques roots in unique changes of the orbital polarization of the electronic states due to distortions of the lattice.
Beyond revealing the key role of hybridized states at the FM/HM interface, our results suggest a clear scheme for generally engineering spin-orbit phenomena. Utilizing the complex interplay of spin and orbital magnetism, spin-orbit coupling, and symmetry, we can tailor the magnitude of SOTs in multilayer devices by designing the orbital polarization of the states near the Fermi energy by strain.


Importantly, our work opens up a route for shaping fundamental spin-orbitronic concepts into competitive technologies by ``dynamically'' tuning the SOTs in perpendicularly magnetized multilayer systems by means of electrically controlled strain. For example, as the strain can be generated locally and imposed on selected parts of the switching area, one can tune the current density such that the DL torque is large enough to switch the magnetization direction in these parts, while it is too small to switch the unstrained parts. In this case it would be possible to switch only selected parts of the area in one run with the given current density. The selected parts can then be altered on demand by utilizing a different configuration of the electric fields, which allows for an additional level of control. Thus, by designing particular strain patterns of the switching area by electric fields, an energy efficient multi-level memory cell capability can be realized, which is practically important, e.g. for the emerging field of neuromorphic computing.\cite{multilevel_AMT}

In addition, we anticipate that strain will not only alter the dynamical properties of topological spin textures but could also modify the Dzyaloshinskii-Moriya interaction\cite{Moriya_1960,Dzyaloshinskii_1957} that may stabilize two-dimensional magnetic solitons. As a consequence, strain offers an efficient means to control the shape and nature of chiral spin structures such as skyrmions\cite{Shibata_2015} and antiskyrmions, which are perceived to hold bright prospects for innovative information processing.


In conclusion we studied the strain response of current-induced SOTs in perpendicularly magnetized W/CoFeB/MgO multilayers grown on a piezoelectric substrate. The SOTs are evaluated by magnetotransport and second-harmonic methods under in-plane strains of different character and magnitude. We find that the strain leads to distinctly different changes of FL and DL torques, with the latter enhanced by roughly a factor of two if a tensile strain is applied parallel to the current flow. Our experimental results are in qualitative agreement with \textit{ab initio} calculations that uncover the microscopic origin of the observed strain effects on SOTs. 
We reveal that the character of strain imprints on the orbital polarization of the electronic states in the ferromagnet, which reflects directly the hybridization with the HM underlayer.
This manifests in a sizable variation of the magnitude of the DL torque while the FL torque remains mostly unaffected. 
The demonstrated possibility to tune the SOTs by means of electric field-induced strain paves a novel path towards to the energy efficient ``dynamical'' control of the current-driven SOT-switching necessary to enable future spintronics applications.

\section*{Methods}
\paragraph{Device fabrication}
The W($5$~nm)/Co$_{20}$Fe$_{60}$B$_{20}$($0.6$~nm)/MgO($2$~nm)/Ta($3$~nm) continuous film was sputter-deposited on top of a bare unpoled two-sides polished piezoelectic commercially available~\cite{mtix} PMN-PT$(011)$ substrate in a Singulus Rotaris commercial sputtering system with a base pressure $<3 \times 10^{-8}$ mbar. The investigated devices were then patterned into Hall crosses with the width of $1 \mu$m by electron beam lithography (EBL) followed by Ar-ion milling of the unnecessary material. The bottom contact of Cr($5$ nm)/Au($50$ nm) [see Fig.~1 (a)] was deposited by DC sputtering in Ar atmosphere. 

Uniaxial in-plane strain was generated by applying an out-of-plane DC electric field across the piezoelectric PMN-PT$(011)$ substrate. The piezoelectric strain response to the applied electric field exhibits a hysteretic behavior.~\cite{Wu_JAP_2011} However, electric fields that exceed the material-specific coercive field pole the substrate and promote a regime where the generated strain is characterized by a linear response to the applied electric field.~\cite{Wu_JAP_2011} 

Before the first measurements, but after the structuring process, the PMN-PT substrate was poled by applying an electric field of $+400$ kV m$^{-1}$. In the following we used in our experiments the DC electric fields that allow us to generate mechanical strain within the linear response regime,~\cite{Wu_JAP_2011} as this provides reliable electrical control over the induced strain.   

\paragraph{2$\omega$ measurements}
The current-induced effective SOT fields presented in this work were measured using 2$\omega$ Hall measurements.~\cite{Hayashi_PRB, Pi_SOT} An AC voltage was applied to the current line and the anomalous Hall voltage was measured with lock-in amplifiers. The in-phase first harmonic ($V_{1\omega}$) and the out-of-phase second harmonic ($V_{2\omega}$) signals were measured simultaneously as a function of in-plane magnetic fields, transverse ($\mu_0H_\mathrm{T}$, along $\pm$ $y$) or parallel ($\mu_0H_\mathrm{L}$, along $\pm$ $x$) to the current flowing in $x$ direction [Fig.~1 (a)].
This allowed us to obtain the transverse ($\mu_0\Delta H_\mathrm{T}$) and the longitudinal ($\mu_0\Delta H_\mathrm{L}$) components of the SOT effective field by using the following expression:~\cite{Hayashi_PRB, Pi_SOT}

\begin{equation}
\mu_0\Delta H_\mathrm{L(T)} = -2 \frac{(B_{x(y)} \pm 2 \xi B_{y(x)})}{1 - 4 \xi^2 },
\label{SOT_eff_fields}
\end{equation}
where $\pm$ sign corresponded to the magnetization direction pointing along $\pm z$, and $B_x\equiv (\frac{\partial V_{2\omega}}{\partial H_x} / \frac{\partial^2 V_{1\omega}}{\partial H^2_x} ) $  and $B_{y}\equiv (\frac{\partial V_{2\omega}}{\partial H_y} / \frac{\partial^2 V_{1\omega}}{\partial H^2_y} ) $. In Eq.~\ref{SOT_eff_fields} $\xi=\frac{R_\mathrm{{PHE}} }{R_\mathrm{{AHE}}}$ is defined as the ratio of the planar Hall effect (PHE) and the anomalous Hall effect (AHE) resistances. We also note that typically negligible $\xi$ is known to be enhanced for W-based multilayers.~\cite{Cho_CAP_2015} Due to extremely large PHE originating from the spin-orbit coupling in W, the ratio can be greater than $1$. In our system $\xi$ was equal to $1.25$ and did not show any dependence on the electric field induced strain.

We measured the SOTs as modified by tensile and compressive strain on the same Hall-cross device, by swapping the current and voltage lines, so that the current direction, i.e. the $x$-axis in Fig.~1 (a), was along $[01\bar1]$ direction of PMN-PT for the tensile strain and along $[100]$ for the compressive strain. 

\paragraph{First-principles calculations}
We performed density functional theory calculations of the electronic structure of a Fe$_{1-x}$Co$_x$/W(001) film as implemented in the full-potential linearized augmented-plane-wave code \texttt{FLEUR}~\cite{fleur}. Exchange and correlation effects were treated within the generalized gradient approximation (GGA),~\cite{Perdew1996} the plane-wave cutoff was $4.1\,$Bohr$^{-1}$, and the muffin-tin radius of magnetic and non-magnetic atoms was $2.42\,$Bohr. In the absence of strain, we assumed the structural parameters as determined in Ref.~\cite{Ferriani2005} The electronic structure of the alloyed compound with variable composition ratio $x$ was treated within the virtual crystal approximation (VCA)~\cite{Bellaiche2000}. Sampling the full two-dimensional Brillouin zone by $24\times 24$ $\mathbf k$-points, we calculated the self-consistent charge density of the perpendicularly magnetized system with spin-orbit coupling. Subsequently, we employed the wave-function information computed on a coarse mesh of $8\times 8$ $\mathbf k$-points for various magnetization directions to generate a single set of so-called higher-dimensional Wannier functions~\cite{Hanke2015}. Based on these functions, we evaluated the linear-response expression of the torkance $\tau_{ij}$~\cite{Freimuth2014a} by using an efficient but accurate generalized Wannier interpolation,~\cite{Hanke2015,Hanke2018} which allows us to access the electronic structure at any $\mathbf k$-point for arbitrary magnetization directions.

\section*{Acknowledgments}
The work was financially supported by the Deutsche Forschungsgemeinschaft (DFG, German Research Foundation) in particular by Grant No. KL1811/18 (318612841) and the Graduate School of Excellence ``Materials Science in Mainz'' (DFG/GSC266). Y.M. acknowledges support by the DFG through the Priority Programm SPP 2137, and additional support was provided by the Collaborative Research Center SFB/TRR 173 (projects A01 - 290396061/TRR173 and B02 290319996/TRR173). J.H. and Y.M. also gratefully acknowledge the J\"ulich Supercomputing Centre and RWTH Aachen University for providing computational resources under project jiff40.



\end{document}